\documentclass[prl,%
 reprint,
 amsmath,amssymb,
 aps,
]{revtex4-2}

\usepackage{graphicx}
\usepackage{dcolumn}
\usepackage{bm}
\begin{document}

\preprint{APS/123-QED}

\title{Cavity-based optical switching via phase modulation in warm rubidium vapor}

\author{G. Booton$^{1}$}

\author{T. Wasawo$^{1}$}%
\author{W.O.C. Davis$^{2}$}%
\author{C. McGarry$^{1,3}$}%
\author{K.R. Rusimova$^{1}$}
\author{A.O.C. Davis$^{1}$}
\author{J. Nunn$^{4}$}
\author{P.J. Mosley$^{1}$}
\affiliation{%
1. Department of Physics, University of Bath, Claverton Down, Bath BA2 7AY, United Kingdom \\
2. School of Mathematical and Physical Sciences, Macquarie University,
Sydney, New South Wales, Australia\\
3. School of Physics, University of Sydney, Sydney, New South Wales, Australia \\
4. ORCA Computing Ltd. 30 Eastbourne Terrace, London W2 6LA UK}
\date{\today}
\begin{abstract}
Optical switching remains a key outstanding challenge for scalable fault-tolerant photonic quantum computing due to the trade-off between speed, bandwidth, and loss. Scalable quantum photonics demands all three, to enable high computational clock rates and resource efficient scaling to large systems. We present a cavity-based optical switch that overcomes this limitation, demonstrating 22 ns rise time, insertion loss of $2.4$ dB, and 17.5 dB extinction ratio. All-optical control is achieved via phase modulation of a signal field detuned from the near-degenerate two-photon absorption ladder in warm rubidium vapor. The ultimate performance of our switch, combining both speed and efficiency, will find applications in active multiplexing, loop-based quantum memory, and feedforward for quantum error-correction protocols.

\end{abstract}

\maketitle

Photonic technologies are at the forefront of the race to leverage quantum advantage \cite{Walmsley:23, quantuminternet}. Active switching for synchronization of probabilistic operations in linear-optical quantum information is essential in scaling up photonic quantum networks \cite{Mendoza:16, PhysRevA.66.053805, PhysRevA.92.053829, Moody_2022, hybridpic, Kaneda:15, Xiong2016}. A practical limitation in this scale-up is the on-demand generation of high purity single photons which requires alleviation by multiplexing in fast, low-loss switch networks or quantum memories \cite{PhysRevA.85.023829, optica_fibermultiplex}. Atomic memory schemes are becoming increasingly more powerful, yet significant challenges still remain in improving storage efficiency and memory lifetime, typically limited to the microsecond scale \cite{PhysRevApplied.18.044058, Tseng:22}. However, with a low-loss optical switch, loop-memory protocols can be implemented with low noise and greater storage efficiencies, without the need for ultra-high vacuum or cryogenic environments, making them a more scalable quantum network architecture \cite{Pang2020HybridMemory, allopqm}. More broadly, low-loss, high-speed optical switches and modulators are critical for enabling dynamic control in quantum key distribution. Low-loss modulators are essential to preserve signal coherence over long distances \cite{Zahidy2024} and for high-bandwidth high-dimensional encoding which enables fast operations \cite{Sit:17,Malik2016}. However, combining high-speed, low-loss operation simultaneously for all-optical switching is still a formidable challenge. 

Commercially available switching technologies are generally optimized for low-loss, high-speed (short rise time) or high-bandwidth (high repetition rate) operation. For example, electro-optic modulators utilizing non-linear crystals such as LiNbO${}_3$, BBO or KDP deliver GHz switching speeds but are often compromised with an insertion loss of greater than 3 dB. Alternatively, Pockels cells offer low-loss operation but with restricted repetition rates that limit the performance \cite{Kaneda:17,8445705}. Micro-electromechanical systems (MEMS) optical switches offer low-loss performance \cite{4840633, Quack2023}, and hold promise for integration with atomic vapor cells where strong interactions between micro-resonator and Rb vapor have been shown \cite{Zektzer:24}. Atomic vapors are a particularly promising avenue for quantum technologies since large phase shifts have been optically induced \cite{Davis_2025} and switching has been shown through the phenomenon of electromagnetically induced transparency (EIT) \cite{PhysRevA.68.041801,PhysRevLett.102.203902}. 

In this letter, we present an all-optical switch that fulfills the set of criteria for quantum applications capable of high-bandwidth operation. Our switch re-routes a signal field through a cavity via a reflective or transmitted port by harnessing a cavity-enhanced light-matter interaction in warm rubidium vapor. Confining the fields to a doubly-resonant ring cavity enhances the interaction of the fields \cite{PhysRevA.96.012338,PhysRevA.76.033804}. As an all-optical atomic system, our switch is inherently capable of operating at high speeds.
In our scheme, we have built a two-port ring cavity where two counter-propagating laser fields interact with atoms in an isotopically pure ${}^{87}$Rb vapor cell. The switching mechanism is provided by modulation of the atomic susceptibility, where a  conditional phase shift imparted on the signal field is induced by a strong control field. 

\begin{figure}[!ht]
   \centering
    \includegraphics[width= 7.5cm]{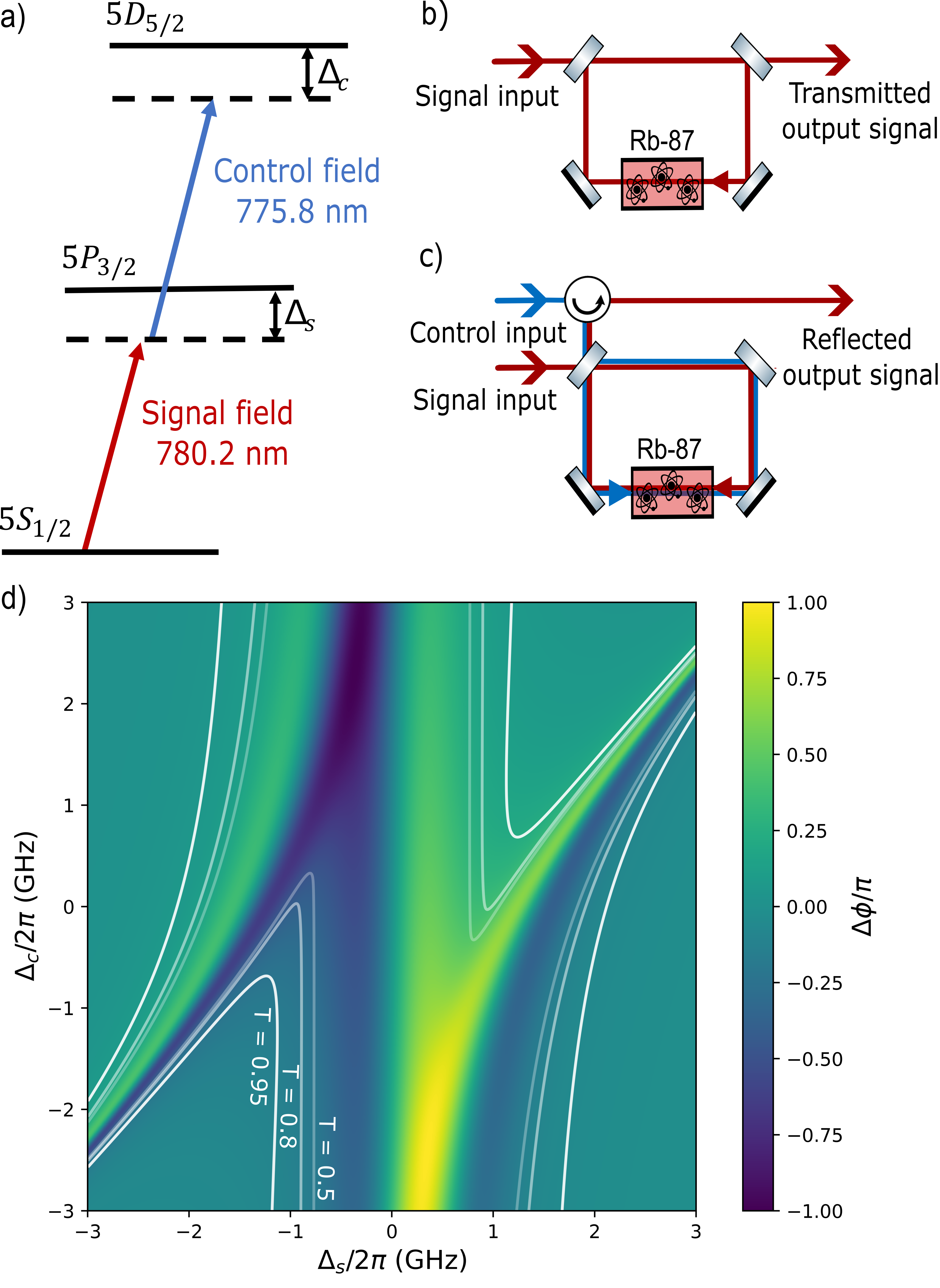}
    \caption{a) The relevant three-level ladder configuration used in ${}^{87}$Rb. b) Switch operation with the control field off. The signal is transmitted through the ring cavity. c) Switch operation with the control field on. A control field excites the two-photon transition and induces a shift in the cavity resonance, hence the incoming signal field is routed to the reflective port. d) 2D theoretical parameter search showing the imparted phase shift, $\Delta\phi$ as a function of both the signal and control field frequency detuning, $\Delta_{s,c}$.  The over-layered (white) contours refers to transmission areas greater than $50\%, 80\%$ and $95\%$ respectively. }
    \label{fig:panel1}
\end{figure}
We take advantage of the near degenerate states in the three-level ladder configuration in ${}^{87}$Rb, as illustrated in \ref{fig:panel1}a). The weak signal field addresses $5S_{1/2}$ to $5P_{3/2}$ transition with a small detuning, $\Delta_s$. A strong control field couples the $5P_{3/2}$ and the $5D_{5/2}$ states with a small detuning, $\Delta_c$. An experimental schematic outlining the switching mechanism is shown in figure \ref{fig:panel1}b) for the control field off. The signal field transmits through the resonant ring cavity with little atomic absorption, hence the signal field is detected on the transmitted output port. The introduction of a control field coupling the higher energy states leads to dressed states and a two-photon resonance, therefore inducing a change in complex susceptibility, $\chi$. This change in susceptibility induced upon the signal field alters the cavity resonance, hence the input signal field is re-routed such that all the signal is detected on the reflected output, as shown in figure \ref{fig:panel1}c). From the complex refractive index, $ n = \sqrt{ 1 + \chi}$, we express the absorption coefficient $\alpha = -2 \kappa \Im(n)$, where $\kappa$ is the wave vector, and the accumulated phase $\phi = \kappa L \Re(n)$ over a distance $L$ of the atomic vapor. The expression for dispersion,  gives rise to the phase shift, $\Delta \phi$, imparted on the signal field, which is the phase difference in the cases of the control field on and off. The accumulated phase shift as a function of both signal and control field frequency detuning, $\Delta_{s,c}$ is shown in figure \ref{fig:panel1}d). We aim for a parameter space that has a high signal transmission through the vapor cell with some phase shift imparted on the signal field. The accumulated phase is cavity enhanced so a high finesse is preferable to increase the imparted phase shift. However, there is a trade-off between finesse and switching speed of our ring cavity. The maximum operational speed is governed by the cavity ring-down time which is estimated down to approximately 20 ns. 


\begin{figure*}[!ht]
    \centering
    \includegraphics[width=15cm]{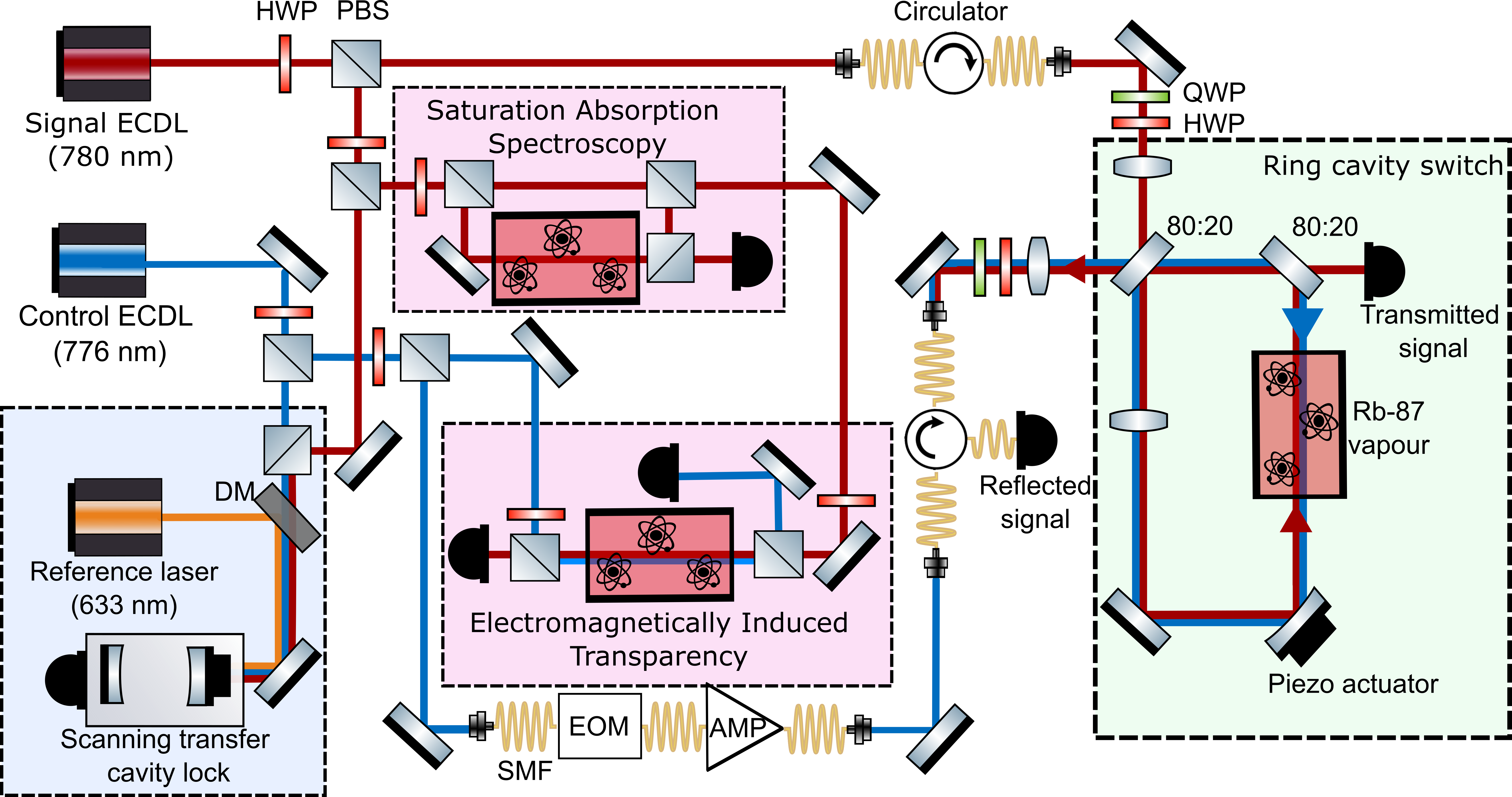}
    \caption{The signal and control fields propagate through saturation absorption spectroscopy and electromagnetic induced transparency experiments as an initial frequency reference. A scanning transfer cavity lock is used to lock the laser frequencies to take experimental data. The signal and control fields then exit single mode fibre and counter propagate around a doubly-resonant ring cavity. Detectors are place on the reflected and transmitted ports. HWP: half wave plate, QWP: quarter wave plate, PBS: polarizing beamsplitter, EOM: electro-optic modulation, AMP: tapered optical amplifier, SMF: single-mode fibre. }
    \label{fig:experiment}
\end{figure*}
The experimental switching setup is shown in figure \ref{fig:experiment}.
The signal field was generated by a 780 nm external cavity diode laser (ECDL, MOGLabs CEL), which was blue detuned from the $5S_{1/2}(F'=2)$ feature. The control field was generated by a 776 nm ECDL (MOGLabs CEL) and was intensity modulated by an electro-optic modulator (Exail NIR-MPX800), and then amplified (MOGLabs MOAL) to produce a square control pulse of approximately 25 mW peak power and down to 42 ns pulse duration. Each laser was frequency stabilized using a scanning transfer cavity \cite{redpitaya}, where the laser fields are superimposed with a frequency stabilized reference laser into a Fabry-P\'erot cavity. Resonances occur as the scanning transfer cavity length is swept as a function of input lasing frequency, and the signal and control field resonances occur within one free spectral range of the reference laser. The relative position of the signal and control field resonances are measured and then an error signal was generated from a RedPitaya locking system.
The optical ring cavity consists of two beamsplitters that have 80:20 reflectivity:transmission ratio and two dielectric mirrors, one of which is fixed and the other mounted onto a piezo-electric transducer for fine cavity length adjustment. The cavity finesse was measured as 18 and a bandwidth estimated from the finesse to be approximately 40 MHz. The cavity ring-up time, $\tau$, can be calculated from the experimentally measured finesse by $\tau = \frac{ \mathcal{F} \cdot L}{c}$, to be 20 ns. This limits the maximum operational frequency of a cavity based switch of our given finesse and size. The atomic medium is isotropically enriched rubidium-87 
vapor, and is placed inside the ring cavity. The vapor is contained within a 5 cm glass cell with anti-reflection coatings, heated to approximately 60$^\circ$C to increase the optical depth whilst minimizing atomic absorption. A lens inside the cavity focuses the signal and control to the centre of the atomic vapor cell. The signal and control light is coupled into the fundamental transverse cavity mode, such that sharp resonances are seen on an oscilloscope. When the control field is applied, the signal field undergoes a $\pi$ phase shift, and is switched from the transmitted to reflected port. The reflected signal travels back on the control arm of the experiment and is detected after a circulator, and the transmitted light exits the cavity on the second port and is detected in free-space.


\begin{figure}[ht!]
    \centering
    \includegraphics[width =8.5cm]{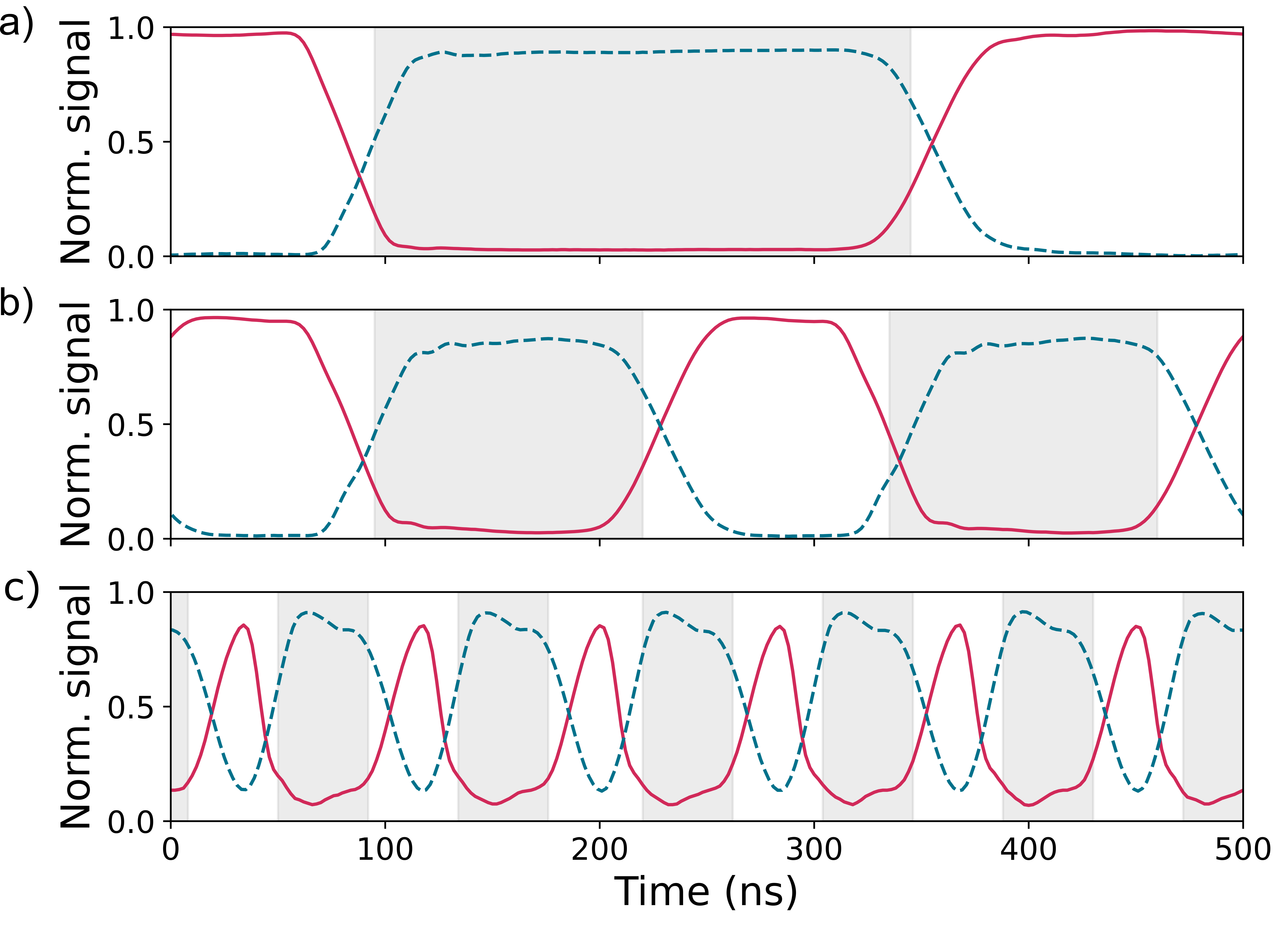}
    \caption{Switching results highlighting three different control modulation rates. The solid red trace shows the transmitted data, the dashed blue shows the reflection data and the shaded regions indicate the control field on. The signal and control field detuning was $-0.65$ GHz. a) 2 MHz control field modulation, with a switching contrast of $89 \%$ and an extinction ratio of 17.5 dB. b) 4 MHz control field modulation, with a switching contrast of $89 \%$ and an extinction ratio of 17.4 dB. c) 12 MHz control field modulation with a switching contrast of $82 \%$ and an extinction ratio of 9.9 dB. }
    \label{fig:results}
\end{figure}

Figure \ref{fig:results} shows the performance from our cavity integrated switch with full $\pi$ phase shifts at three different modulation rates. The transmission is shown by the solid red line, and reflection by the blue dashed line. The shaded regions indicate the control field on. The transmission with the control field on,  $T_{\mathrm{on}}$, is the average value of the red trace in the shaded region, and similarly for the $T_{\mathrm{off}}$ in the unshaded regions. The switching contrast, $C$ on the transmitted signal is defined as, 
\begin{equation}
    C = \frac{T_{\mathrm{on}}- T_{\mathrm{off}} }{ T_{\mathrm{on}}+ T_{\mathrm{off}}}.
\end{equation}
A switching contrast of $89 \%$ is achieved with a extinction ratio of 17 dB in figure \ref{fig:results}a) and b). The limit of the switch speed is experimentally demonstrated in figure \ref{fig:results}c), with an open eye diagram at 42 ns switching time with a switching contrast of $82\%$ and a rise time of 22 ns. The data supports the predicted cavity ring-up time and therefore the fundamental limit of our experiment. 

\begin{figure}[h!]
    \centering
    \includegraphics[width=8.5cm]{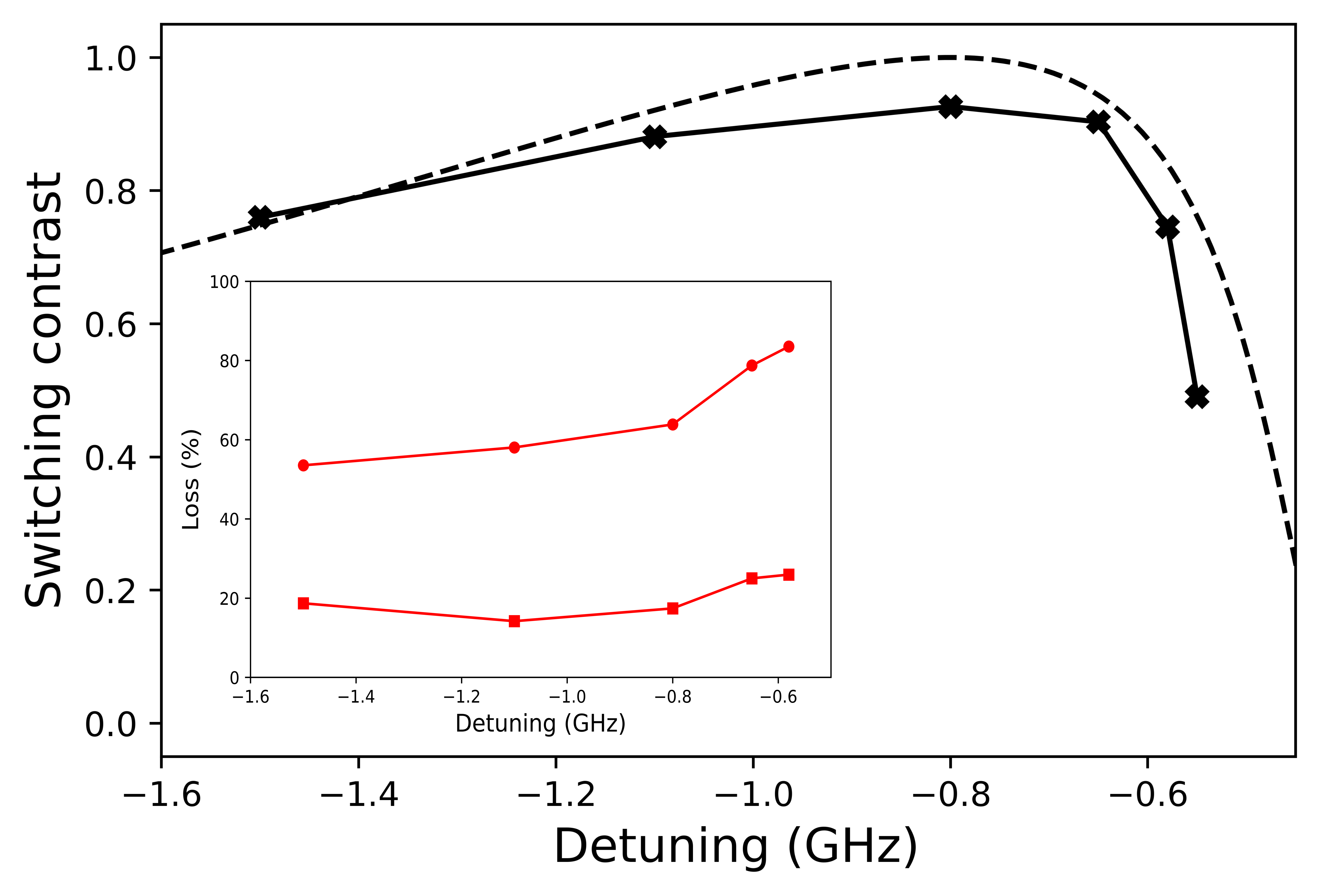}
    \caption{ Systematic characterization of the switching contrast. The signal and control frequency detunings are stepped through simultaneously such that $\Delta_c = \Delta_s$. The experimental data (solid black) is shown alongside simulation data (dashed black) of the switching contrast. The inset shows measurements of the insertion (red circle) and intra-cavity (red square) loss.}
    \label{fig:combined sys}
\end{figure}
The switch performance in terms of contrast and loss as a function of the simultaneous signal and control field detuning is shown in figure \ref{fig:combined sys}, where the switching contrast is plotted with experimental data alongside a analytic model. The model considers the transmission of the signal field through an optical cavity with an additional phase accumulated on the signal from the rubidium vapor. The transmission function is derived from the coupled ring-cavity dynamics which relates the incoming signal field directly to the transmitted output, $T(\phi)$, with a phase $\phi$:
\begin{equation}
    T(\phi) =  \frac{t_1^2 t_2^2 \eta }{1 + r_1^2 r_2^2 \eta - 2 r_1 r_2 \eta \cos(\phi)}\,,
    \label{eq:T}
\end{equation}
where $r_1$ and $r_2$ are the beamsplitter reflectivities, and $t_1$ and $t_2$ are the transmission of the beamsplitters, and $\eta$ is the cavity loss.
The phase in equation \ref{eq:T} is then updated with the refractive index modulation induced from the atomic susceptibility such that,
\begin{equation}
    \phi = \frac{2 \pi}{\lambda_s }[ ( \Re(n) -1)L_v +L_c)]\,,
\end{equation}
where $L_v$ and $L_c$ are the length of the atomic vapor cell and the cavity length respectively. The modeled switching contrast of the transmission is plotted when stepped through the signal and control frequency detuning, under the condition where $\Delta_c = \Delta_s$ throughout.
Two free parameters, temperature and intra-cavity power, were determined by a least squares optimization between the experimental and simulated data. The best fit corresponded to a vapor cell temperature of 59$^\circ$C and intra-cavity control power of 500 mW. These values are close to the experimental values where the temperature of the vapor cell was measured externally to be approximately 60$^\circ$C and the intra-cavity power is approximated by the input control power $P = 25$mW with cavity enhancement $P \cdot \mathcal{F}$. 
The insertion loss and the intra-cavity loss are plotted in the inset of figure \ref{fig:combined sys}. 
The insertion loss is defined as the ratio of the transmitted signal $T_\mathrm{on}$ to the input signal. The intra-cavity loss represents the loss induced from the cavity, and is defined as $1 -(T_\mathrm{on} + R_\mathrm{on})$. The cavity loss at optimum operation is approximately $ 17 \%$. The intra-cavity loss is determined to come from scattering on internal optics, imperfect anti-reflection coatings on the vapor cell and rubidium condensation on the windows of the cell inside the optical cavity. 
At frequencies close to atomic resonance, a poor switching contrast is seen due to increased atomic absorption, leading to high insertion loss, however low intra-cavity loss is observed due to observation of EIT. The contrast improves with further detuning due to less atomic absorption, and the signal transmission increases. The switching contrast begins to tail off at far-detuned frequencies due to more control power being needed to provide sufficient phase shift, whilst the loss of the system remains relatively low. 

In conclusion, we have presented an all-optical switch that has the ability to operate at high speeds, with a rise time of 22 ns. The rise time is limited by the cavity ring down time, theoretically estimated as 20 ns and is validated by the experimental limit shown in this paper. The insertion loss of the device is $3.7$ dB at optimum operation, which can be reduced to $2.4$ dB at far-detuned laser frequencies, which is accompanied by a slight reduction in the switching contrast. The insertion loss is also further reduced to $1.6$ dB with a reduction in the mirror reflectivity, but this results in a smaller extinction ratio of our switch. The ratio between the experimentally measured cavity finesse and the maximum theoretical finesse is 90\%, meaning that intra-cavity loss is minimal. Intra-cavity loss is measured at approximately $17 \%$, predominately due to loss on the internal optics. Higher cavity finesse can be achieved by miniaturization of the cavity, which would significantly reduce insertion loss due to increased coupling and a reduction of intra-cavity optics and allow for faster switching speeds due to a reduced cavity mode volume. Micro-structured optical fibre is one route to achieve this, and an all-fibre integration of our switch would provide seamless integration into existing fibre networks \cite{Camerons_perspective}.   
The all-optical switching in rubidium vapor presented here demonstrates a marked improvement over existing optical switching technology, and therefore is a significant step forward towards scalable quantum technologies.

\begin{acknowledgments}

\textit{Acknowledgments}-This work is supported by the Air Force Office of Scientific Research (AFOSR) under award number FA8655-22-1-7024. We acknowledge funding from Innovate UK project \textit{The Quantum Data Centre of the Future} (application number 10004793) and The UK National Quantum Technologies program through the Hub in Quantum computing and simulation (EPSRC grant EP/T001062/1) and the Hub for Quantum Computing via Integrated and Interconnected Implementations (EPSRC grant EP/Z53318X/1).

\end{acknowledgments}

\nocite{*}

\providecommand{\noopsort}[1]{}\providecommand{\singleletter}[1]{#1}%

\end{document}